\documentclass[iop]{emulateapj}
\usepackage{hyperref}

\slugcomment{Accepted for publication in The Astrophysical Journal}

\shorttitle{The Progression of Star Formation in the RMC}
\shortauthors{Ybarra et al.}

\begin{document}


\title{The Progression of Star Formation in the Rosette Molecular Cloud}


\author{Jason E. Ybarra\altaffilmark{a}}
\affil{Department of Astronomy, University of Florida, Gainesville, FL 32605, USA}
\email{jybarra@astro.ufl.edu}
\altaffiltext{a}{NASA Goddard Space Flight Center GSRP fellow}

\author{Elizabeth A. Lada}
\affil{Department of Astronomy, University of Florida, Gainesville, FL 32605, USA}

\author{Carlos G. Rom{\'a}n-Z{\'u}{\~n}iga}
\affil{Instituto de Astronom{\'i}a,  Universidad Nacional Aut{\'o}noma de Mexico,
Unidad Acad{\'e}mica de Ensenada, Apdo. Postal 22860, Ensenada, B. C., Mexico 
}

\author{Zoltan Balog}
\affil{Max-Planck-Institut f\"ur Astronomie, Heidelberg, Germany}

\author{Junfeng Wang}
\affil{Department of Physics and Astronomy \& Center for Interdisciplinary Exploration and Research in Astrophysics (CIERA), Northwestern University, 2131 Tech Dr, Evanston, IL 60208}

\author{Eric D. Feigelson}
\affil{The Pennsylvania State University, University Park, PA 16802, USA}



\begin{abstract}

Using {\it Spitzer Space Telescope} and {\it Chandra X-ray Observatory} data, we identify YSOs in the Rosette Molecular Cloud (RMC).
By being able to select cluster members and classify them into YSO types, we are able to track the progression of star formation locally within the cluster environments and globally within the cloud.
We employ nearest neighbor method (NNM) analysis to explore the density structure of the clusters and YSO ratio mapping to study age progressions  in the cloud.
We find a relationship between the YSO ratios and extinction which suggests star formation occurs preferentially in the densest parts of the cloud and that the column density of gas rapidly decreases as the region evolves. This suggests rapid removal of gas may account for the low star formation
efficiencies observed in molecular clouds.
We find that the overall age spread across the RMC is small. Our analysis suggests that star formation started throughout the complex around the same time. Age gradients in the cloud appear to be localized and any effect the \ion{H}{2} region has on the star formation history is secondary to that of the primordial collapse of the cloud.

\end{abstract}





\section{Introduction}

Determining the effect environment has on the star forming process is imperative to gaining
a full understanding of star formation. Most stars form in embedded clusters distributed throughout a molecular cloud \citep{ladalada2003}. These embedded clusters are often subject to different environmental conditions based on physical their location in the cloud. 

The Rosette Molecular Cloud (RMC) has a ideal layout for studying star formation and the effects of environment. At its northeastern end lies the Rosette Nebula which contains the OB association NGC 2244. 
The RMC contains many embedded clusters throughout the length of the cloud.
The near-infrared study of \citet{phelps1997} identified 7 embedded clusters (PL01-07) within the cloud.  
Subsequent mid-infrared, near-infrared, and X-ray studies have refined and added to this list.
\citet{rom2008} confirmed the NIR clusters and also discovered two more (REFL08-REFL09) using the density distributions of near-infrared excess (NIRX) sources. 
Within the PL04 region, \citet{rom2008} found the peak of NIRX sources (PL04a) to be spatially coincident with the NIR nebulosity.  
\citet{poulton2008} found a concentration of {\it Spitzer} identified YSOs (cluster D = PL03b) offset by 4.3 arcmin (2.0 pc) from the center of the NIRX distribution PL03a. 
Additionally, a distribution of YSOs (cluster C = PL02b) was found just south of the NIR cluster PL02. 
\citet{wang2009}
found a cluster of X-ray sources in the northern end of PL04 indicating the
presence of a population of Class III sources. The center of this distribution (cluster XC = PL04b) is offset from the peak of the NIRX distribution by 3.3 arcmin (1.5 pc). 
Additionally, a small distribution of X-ray sources (XB2) was found spatially coincident with MIR cluster PL02b. 

These embedded clusters are found in different environments.
Clusters near the OB association are within the ionization front of the \ion{H}{2} region. Further away from the ionization front, is a region that is associated with a high concentration of gas and star formation, where it is estimated almost half of the star formation in the cloud is taking place  \citep{rom2008}. This region may have experienced a shock front passing through it. 
Finally at the back of the cloud is a region where two embedded clusters have formed beyond the influence of the nebula.

\citet{rom2008} studied the RMC with a deep near-infrared {\it JHK} survey. In their analysis of NIR excess sources they found that the excess fraction within the clusters increased with distance away from the the center of the Rosette Nebula. This suggested an underlying age sequence in the cloud where cluster age decreases with increasing distance from the nebula. 
This sequence extends beyond the influence of the \ion{H}{2} region and it was suggested that its origin is from the formation and evolution of the cloud.

In this study we take a closer look at the distributions of these young sources and focus on tracking the progression of star formation within the cloud.

\section{Observations}

This study makes use of data obtained from the {\it Spitzer Space Telescope}, the {\it Chandra X-ray Observatory}, and the FLAMINGOS instrument on the KPNO 2.1 m telescope. Figure 1 is a {\it Spitzer} IRAC 3-color image of the RMC survey region. The blue dashed lines show the extent of the FLAMINGOS {\it JHK} survey and the red lines show the boundaries of the {\it Chandra} x-ray observations. 
The locations and names of the embedded clusters are shown.

\subsection{Spitzer IRAC observations and data reduction}

For this study we use IRAC 3.6-8.0 $\mu$m and MIPS 24 $\mu$m data from program 3394
(PI: Bonnel) available in the {\it Spitzer } archive. 
The IRAC mapping covers a total projected area of 1.4 $\times$ 0.8
degrees$^2$ for all four IRAC bands. 
Each IRAC pointing has a field of view of 5.12\arcmin $\times$ 5.12\arcmin\ with 3 dither positions and an exposure time of 12 seconds per frame. 
Additionally, we observed the NIR cluster REFL09 with IRAC from a separate program 40359 (PI: Rieke). 
The REFL09 IRAC observation ($\alpha$ = 06:35:09.11 $\delta$ = +03:41:13.7) is a single pointing with a field of view of 5.12\arcmin $\times$ 5.12\arcmin, 3 dither positions, and an exposure time of 12 seconds per frame. 

The IRAC frames were processed using the Spitzer Science Center (SSC) IRAC Pipeline
v14.0, and mosaics were created from the basic calibrated data (BCD)
frames using a custom IDL program (see \citet{Guter08} for details).
The MIPS frames were processed using the MIPS Data Analysis Tool \citep{Gordon2005}.
Source detection and aperture photometry was perform using the IDL software package PhotVis 
which is based in part on DAOPHOT 
\citep{gutermuth2004, Guter08}. 
The MIPS 24 $\mu$m frames were processed using the MIPS Data Analysis Tool. 
PSF photometry was performed on the MIPS data using the DAOPHOT IRAF package. 

\notetoeditor{Please print Fig 1 across both columns}
\begin{figure*}
\plotone{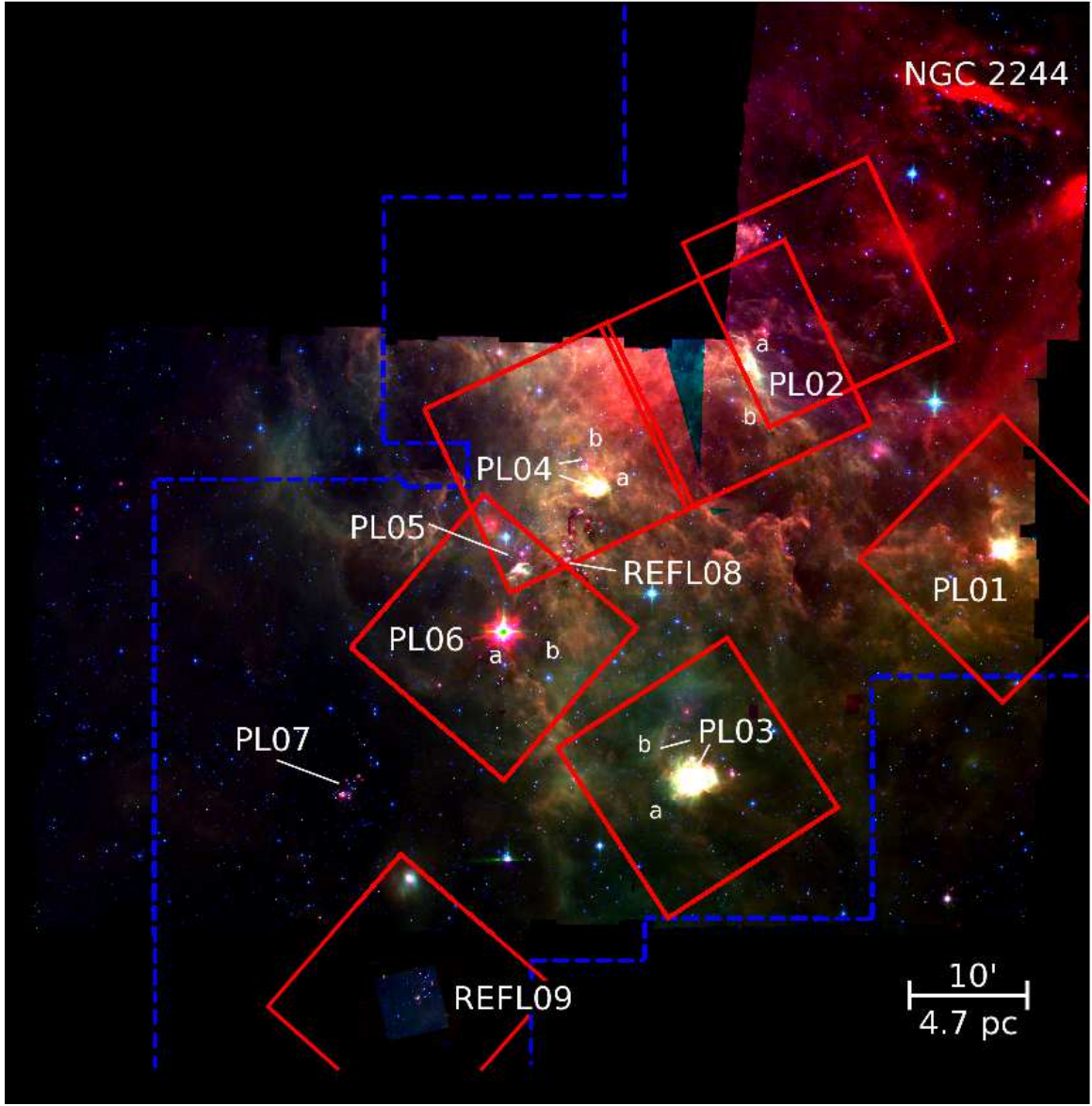}
\caption{
{\it Spitzer} 3-color image of the Rosette Molecular Cloud
[4.5 $\mu$m (blue), 5.8 $\mu$m (green), 24 $\mu$m (red)].
Blue dashed lines show the boundaries of the FLAMINGOS {\it JHK} survey \citep{rom2008}.
Red lines show the boundaries of the {\it Chandra} x-ray observations \citep[this paper]{wang2009}.
}
\end{figure*}

\subsection{Chandra observations and data reduction}

The embedded Rosette clusters studied in this paper were observed with
the Imaging Array of the {\em Chandra} Advanced CCD Imaging
Spectrometer (ACIS-I; \citet{garmire2003}) onboard of the {\em Chandra
  X-ray Observatory}, which has a $17\arcmin \times 17\arcmin$ field
of view in a single pointing.  
These observations were taken on 2010 December 3 (ObsID 12388, covering the PL3 cluster), 2010 December 10
(ObsID 12142, PL6), 2011 January 14 (ObsID 12387, PL1), and 2011
January 18 (ObsID 12386, REFL 9), with net exposure time of 19.6 ks,
39.3 ks, 24.5 ks, and 34.6 ks, respectively.  
They significantly supplement the previous {\em Chandra} campaign of the Rosette complex
reported in \citet{townsley2003} and 
\citet{wang2008,wang2009,wang2010}
which consisted of four $\sim$20 ks ACIS-I snapshots in January 2001, a deep
75 ks ACIS-I image in January 2004 centered on the O5 star HD~46150 in
NGC 2244, and one 20 ks ACIS-I pointing at the NGC 2237 sub-cluster in
2007.  
All images were taken in ``Timed Event, Very Faint'' mode (5 pixel $\times$ 5 pixel event islands).

We follow the same customized data reduction described in \citet{wang2007,wang2008}
using the Chandra Interactive Analysis of
Observations (CIAO, \citet{fruscione2006}; version 4.3) package
provided by the Chandra X-ray Center.  
A detailed description of the X-ray data analysis will be presented in a separate paper (Wang et
al. 2013, in preparation).  
For each of the ACIS fields, we extracted X-ray images in the 0.5--7~keV band, and applied the source detection
algorithm {\it
  wavdetect}\footnote{{\url{http://cxc.harvard.edu/ciao/threads/wavdetect/}}}
  \citep{Freeman02} with a range of wavelet scales (from 1 to 16
  pixels in steps of $\sqrt 2$) and a source significance threshold of
  $1 \times 10^{-5}$ to produce a list of candidate sources.
X-ray event extraction was made with our customized IDL script {\it
  ACIS Extract}\footnote{\url{http://www.astro.psu.edu/xray/docs/TARA/ae\_users\_guide.html}}
  \citep[{\it AE};][]{Broos10}.  
Using the {\it AE}-calculated probability $P_B$ that the extracted
events are solely due to Poisson fluctuations in the local background,
we rejected sources with $P_B > 0.01$, i.e. those with a 1\% or higher
likelihood of being a background fluctuation. 
The trimmed source list includes 431 valid X-ray sources.

\subsection{Near-infrared photometry data}

This study makes use of photometry data from the
FLAMINGOS {\it JHK} imaging survey of the RMC \citep{rom2008}
which is part of the NOAO survey program "Toward a Complete Near-Infrared
Spectroscopic and Imaging Survey of Giant Molecular Clouds" (PI: E. A. Lada). 
The completeness limits for the survey are $J=17.25$, $H=18.00$, and $K=18.50$. 

\section{Analysis}

\subsection{Dust extinction distribution in the RMC}  

A dust extinction map was made with the NICEST algorithm
\citep{lombardi2009} on the FLAMINGOS photometry catalog. 
We removed Class 0/I and Class II sources from the catalog as those sources have
large intrinsic red colors and tend to bias the map near the centers of clusters.
The extinction is estimated toward background sources by comparing the 
source NIR colors with the intrinsic color distribution measured from a
 nearby extinction-free control region.
The control field is located at a similar galactic latitude as the RMC and was selected from an IRAS 25 $\mu$m map as being devoid of dust thermal emission 
\citep[see][Fig.2]{rom2008}.
Spatial smoothing is then applied to the extinction values to create the extinction map. 
The smoothing creates a map of the weighted mean as a function of position on the sky. 
The weighting function is composed of a Gaussian which weighs the individual extinction measurements based on angular distance from the center of the map point plus a correction term which compensates for the bias due to the low numbers of background sources at high extinctions.
In the case of the RMC, the optimal 
Gaussian width was found to be 90$''$. 
This is capable of resolving column density structures of the molecular cloud
with projected sizes of about 0.5--1.0 pc.

Figure 2 shows the extinction map of the RMC. 
The contour levels indicate A$_{\rm V} = 8, 10, 12, 14, 16, 20$ mag. 
The locations of the embedded clusters are indicated. 
Comparison of the location of the embedded clusters with the extinction distribution 
suggest a correspondence between star formation with the highest extinction regions of the cloud.

\notetoeditor{Please print Fig 2 across both columns}
\begin{figure}
\plotone{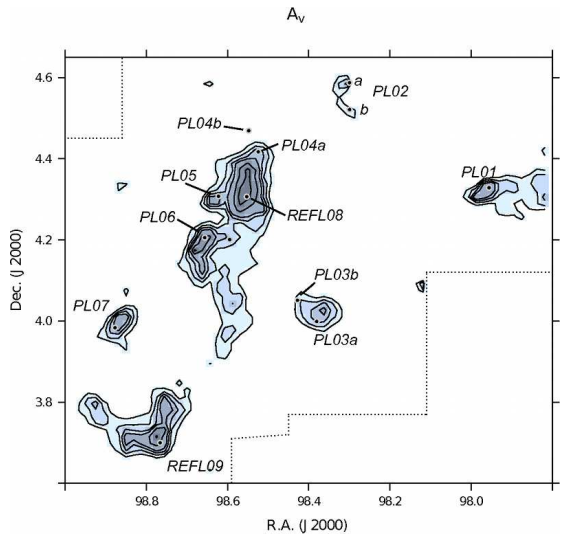}
\caption{NICEST extinction map of the Rosette Molecular Cloud. The contours 
represent extinction levels A$_{\rm V}$ = 8,10,12,14,16,20 mag. 
Location of embedded clusters are indicated.
}
\end{figure}

\subsection{Identification of YSOs}    

{\it Spitzer} and {\it Chandra} ACIS data were used to identify 
Young Stellar Objects (YSOs) in the Rosette Molecular Cloud.
In order to select the Class I/0 and Class II sources, we employ the color cuts of 
\citet{Guter08} and \citet{kryukova2012} to our {\it Spitzer} catalog. 
For objects which have IRAC (3.6 $\mu$m, 4.5 $\mu$m, 5.8 $\mu$m, 8.0 $\mu$m) and MIPS 24 $\mu$m detections, we identify Class I/0 sources 
using the following criteria: 
\begin{eqnarray}
& [4.5]-[5.8]  > 1 & \nonumber\\
& \mbox{or} & \nonumber\\
& [4.5]-[5.8] > 0.7 \mbox{~and~} [3.6] - [4.5]  > 0.7 & \nonumber
\end{eqnarray}
and
\begin{displaymath}
[4.5] - [24] > 4.7
\end{displaymath}
From the remaining catalog we select Class II sources that have 
all of the following criteria:
\begin{eqnarray}
& [4.5]-[8.0] > 0.5 & \nonumber\\
& [3.6]-[5.8] > 0.35 & \nonumber\\
& [3.6]-[5.8] \leq 3.5([4.5]-[8.0]) - 1.75 & \nonumber
\end{eqnarray}
The Gutermuth [4.5]--[5.8] color cut that distinguishes between Class I/0 and Class II sources is particularly useful as the extinction curve is relatively flat 
in that wavelength range and thus the color cut is insensitive to extinction. 
For sources without an IRAC 4.5 $\mu$m detection, we select Class I/0 sources by
\begin{displaymath}
[5.8] - [24] > 4.5 \mbox{~and~} [24] < 6
\end{displaymath}
and for Class II sources
\begin{displaymath}
[3.6] - [5.8] < 0.35 \mbox{~and~} 2.0 < [5.8]-[24] \leq 4.5
\end{displaymath}

For objects that do not have a MIPS 24 $\mu$m detection, we use the previous IRAC color cuts for objects with a [4.5]--[5.8] color and the additional requirement of
\begin{displaymath}
[5.8]-[8.0] < 1
\end{displaymath}
in order to filter out AGN and PAH galaxies.

Class III sources do not display significant infrared excess and thus need to be identified another way. Fortunately, YSOs are known to emit X-rays at levels that can range  many orders of magnitude above main sequence stars \citep{preibisch2005, feigelson2007}. Thus, X-ray observations can efficiently identify YSOs in molecular clouds. 
Class III candidate sources were selected from our {\it Chandra} ACIS observations and the previously published X-ray catalog of \citet{wang2009}.
Sources with colors of Class I/0 or Class II objects were then removed to create a catalog of Class III sources.
In order to deal with extragalactic contaminants in our sources list, we also removed sources that 
did not have a NIR counterpart in the FLAMINGOS catalog.
In our analysis of the X-ray properties, we found that the sources without a NIR counterpart had an average hardness ratio consistent with extragalactic sources (Wang et al., in preparation).

\subsection{Spatial distribution of the different YSO classes} 

By analyzing the distributions of YSO classes separately one can probe
the evolution of star forming regions. 
The different YSO classes represent different evolutionary stages, with Class 0 and I sources representing the youngest sources still embedded in their envelopes, Class II sources are a later stage of sources still accreting material from their disks, and Class III sources are diskless pre-main sequence stars. 

We employed the k-nearest neighbor (k-NN) density estimation
algorithm, often referred to as the nearest neighbor method (NNM), to
analyze the structures and distributions of the YSO classes.  
This method allows us to map the density distributions and subsequently identify regions of clustering \citep{casertano1985}.
The algorithm measures the distance, $D_{k}$, between a source point and its k$^{\rm{th}}$  nearest neighbor and estimates the local stellar number density as $\mu_{k} = (k-1)/\pi D_{k}^{2}$.
For the Class I/0 objects we used a k=5, and for the Class II and Class III sources we used k=10.
Using k values larger than one has the advantage of lessening the
influence possible non-members of the set, e.g. background galaxies,
have on the local density estimate \citep{ferreira2010}. 
This is important when using {\it Spitzer} data as many background galaxies have colors that are similar to those of YSOs in both the near- and mid-infrared \citep{foster2008}. 

Figure 3 presents the NNM density maps of the Class I/0, Class II, and Class III sources in the RMC. 
The dashed lines show the spatial boundaries of the survey data used to identify the sources.
The embedded clusters are spatially coincident with the high stellar density regions in the maps.
Using the NNM maps of the Class II and Class III sources, we have identified a new cluster (PL06b).
The maps reveal that the different YSO classes can have very different spatial distributions. 
For example the Class I/0 and Class III concentrations appear to be separate while the Class II sources appear more wide spread.
This is especially evident in the central region of the cloud. 
Figure 4 shows the NNM density maps of the central region for the Class I/0 and Class III sources. 
A high density of Class I/0 sources is found in the center of this region and is spatially coincident with the highest extinction.
In contrast the older Class III sources are found primarily around the perimeter on north and east sides of the central region. 
Class II sources are found throughout the central region. 
By numbers, the Class II sources dominate the cloud.
The different density distribution of the YSO classes in the central region suggest a
progression of star formation from the outer north and east sections towards the central.

\begin{figure}
\includegraphics[width=0.47\textwidth]{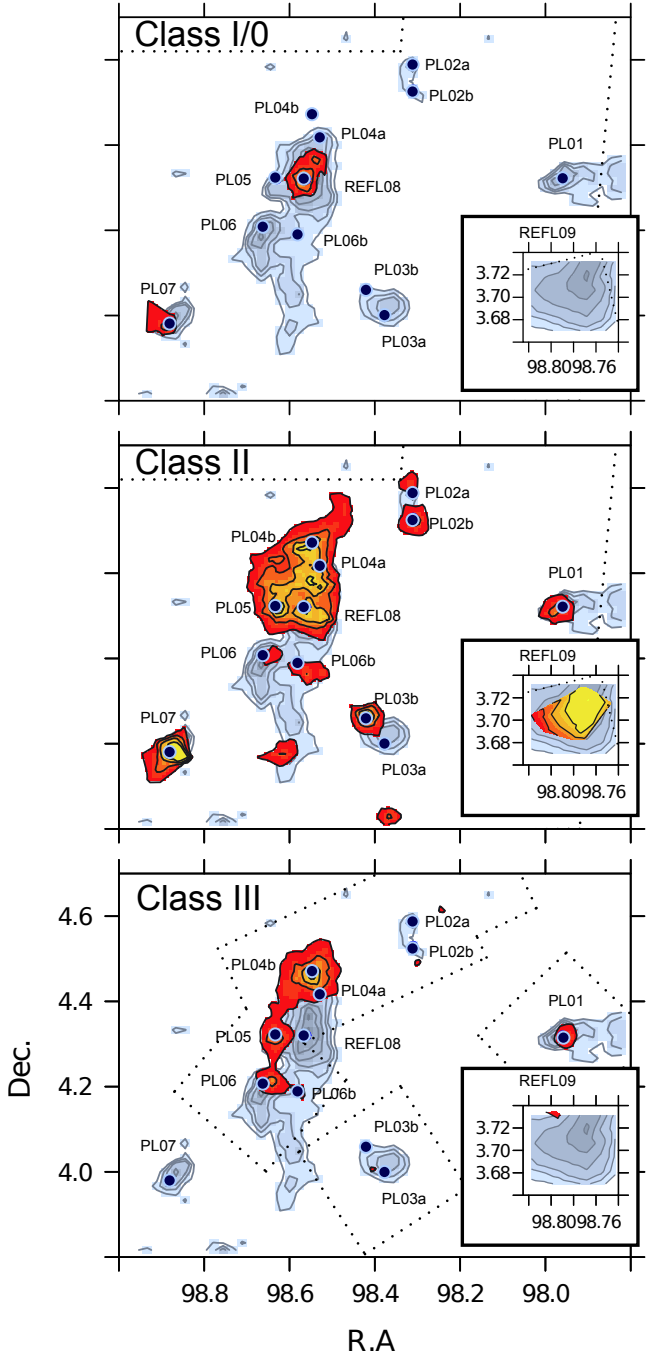}
\caption{Nearest-Neighbor (NNM) density maps of the RMC for Class I/0, Class II, and Class III
sources.
The density contours are $\mu = 2.6, 4.6, 7.7, 12.8$ stars pc$^{-2}$.
The background contours are $A_{V} = 8,10,12,14,16,20$.
Dotted lines in the first two panels show the coverage of the {\it Spitzer} observations and the
dotted lines in the last panel show the coverage of the {\it Chandra} observations.
}
\end{figure}

\begin{figure}
\plotone{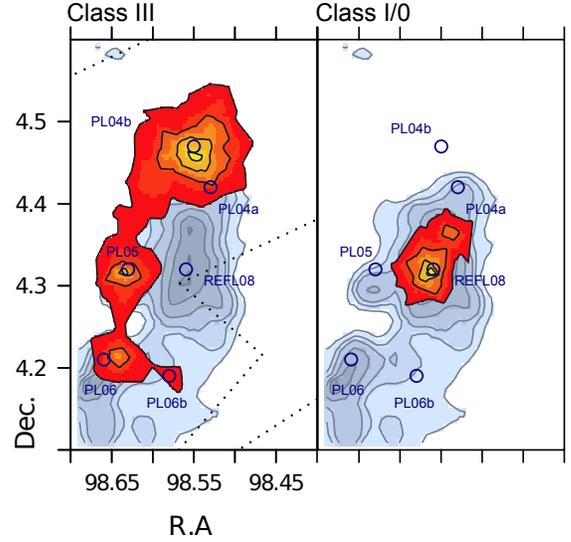}
\caption{Nearest-Neighbor (NNM) density maps of the RMC main core region
 for Class I/0 and Class III sources.
The density contours are $\mu = 2.6, 4.6, 7.7, 12.8$ stars pc$^{-2}$.
The background contours are $A_{V} = 8,10,12,14,16,20$}.
\end{figure}

\subsection{Age distribution through YSO ratios}    

In order to study the progression of star formation in the RMC, we investigate how the ratios between different YSO classes change
throughout the cloud.
As time progresses, Class I/0 sources will lose their envelopes and evolve into Class II sources. 
Because the lifetime of the Class I/0 phase is less than the Class II phase,
the ratio of Class I/0 to Class II sources should decrease with
the age of the region \citep[cf.][]{myers2012}. 
Similarly, as time progresses, Class II sources will lose their disks
and evolve into Class III sources. 
Disk fractions in clusters have an empirical relationship with age
\citep{haisch2001}. 
From different {\it Spitzer} IRAC studies of young clusters, the
average disk fractions at different cluster ages are 75\% at 1 Myr,
50\% at 2-3 Myr, 20\% at 5 Myr, and 5\% at 10 Myr
\citep{williams2011}. 
Thus, the Class II to Class III ratio within a region will also
decrease with age (Table 1). 
This relationship between the Class II to Class III ratio and age allows us to probe age progressions within the cloud.
\begin{deluxetable}{ccc}
\tablehead{ \colhead{Age (Myr)} & \colhead{D.F.} & \colhead{$R_{II:III}$}
}
\startdata
1 & 75\% & 3.0 \\
2-3 & 50\% & 1.0 \\
5 & 20\% & 0.3
\enddata
\tablecomments{Columns 1 and 2 are cluster ages and average disk fractions from the literature \citep{williams2011}. Column 3 is the associated Class II to Class III ratio.
}
\end{deluxetable}

We constructed ratio maps by estimating the ratio of the number of one class of sources, $N_{1}$, to that of another, $N_{2}$, (e.g. Class I/0 to Class II) within a projected region of the sky across a grid . 
We chose the size of the region by trying to minimize both resolution and uncertainty.
The probability density function of the ratio, $R$, is
\begin{displaymath}
p(R | N_{1},N_{2}) = \frac{R^{N_{1}} (N_{1}+N_{2}+1)!}
{ (R+1)^{N_{1}+N_{2}+2} N_{1}! N_{2}! }
\end{displaymath}
assuming a uniform prior for the ratio \citep{jin2006}.
We use the expectation value of the ratio as our estimator, 
\begin{displaymath}
\hat{R}  = \frac{N_{1}+1}{N_{2}},
\end{displaymath}
with variance,
\begin{displaymath}
\sigma_{R}^{2} = \frac{(N_{1}+1)(N_{1}+N_{2}+1)}
{N_{2}^{2}(N_{2} -1)}
\end{displaymath}
In each point on the grid, for a non-zero ratio value, we require $\sigma_{R}/\hat{R} \leq 0.50$. 
For the Class I/0 to Class II ratio we chose a circular region with radius 1.1 pc and 
for the Class II to Class III ratio a circular region with radius 1.4 pc. 
The radius for the Class II to Class III ratio is larger because the
magnitude cuts (see next paragraph) reduce the number of sources. 
The regions are sampled across the grid at intervals of one-third the region radius. 
The created ratio map provides a visual representation of age gradients and can then map the progression of star formation. 

In order to compare Class II and Class III sources it is necessary to
make sure the samples are uniform and unbiased \citep{gutermuth2004}. 
We first restrict the samples to sources which are detected in the FLAMINGOS {\it JHK} survey. 
We estimate the Class II sample is complete to $H = 15$ and the Class III sample to 
be complete to $H = 14$. 
We use the H-band luminosity as the emission from disks at this wavelength is negligible.
We de-redden our samples to a 2 Myr isochrone \citep{baraffe1998} and limit the samples 
to $H \leq 14$, making the samples complete to the same depth across the cloud.

Figure 5 shows the two ratio maps, the Class II to Class III on the
left and the Class I/0 to Class II on the right. 
These maps are over-plotted on top of the extinction contours. 
For the Class II to Class III ratio, the contour levels are $R_{II:III} = 0.5, 1.0, 2.0, 3.0$. 
For the Class I/0 to Class II ratio, the contour levels are $R_{I:II} = 0.1, 0.3, 0.4, 0.5$.
Visual inspection of the maps suggest a correlation between the ratios and extinction, where 
higher ratio values are spatially correlated with higher extinction.
\notetoeditor{Please print fig across both columns} 
\begin{figure*}
\plotone{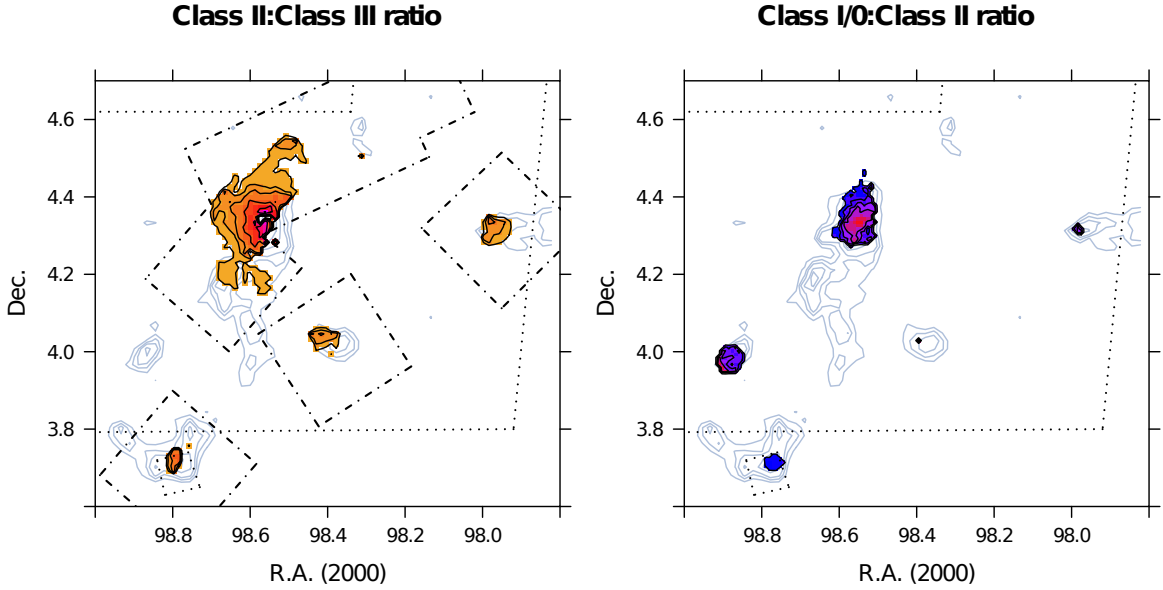}
\caption{Left: Class II to Class III ratio map. The contour levels are $R_{II:III} = 0.5, 1.0, 2.0, 3.0$.
 Right: Class I/0 to Class II ratio map.  The contour levels are $R_{I:II} = 0.1, 0.3, 0.4, 0.5, 0.7$.
The background contours are A$_{\rm V}$ = 8,10,12,14,16. Dotted lines show the coverage of the {\it Spitzer} observations and dash dotted lines show the coverage of the {\it Chandra} observations }
\end{figure*}

To further investigate the correlation between these ratios and extinction we also determine the average extinction in each region.
We binned the regions by extinction into 1 mag bins and calculated the weighted mean ratio in each bin. Figure 6 presents a plot of extinction bin versus mean ratio of Class II to Class III sources.
The figure shows that the ratio increases monotonically with extinction. 
The plot can be fitted with a shallow linear fit for A$_{\rm V} < 13$ mag, 
and a steeper linear fit for A$_{\rm V} > 13$ mag. 
Thus, decreasing age is related to increasing extinction.
\begin{figure}
\plotone{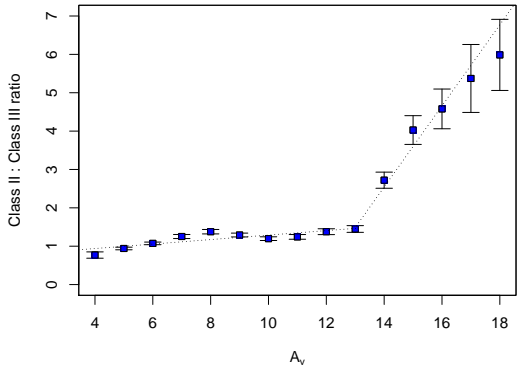}
\caption{ Plot of extinction bin versus mean Class II to Class III ratio. 
}
\end{figure}
Figure 7 shows the extinction bin versus mean Class I/0 to Class II ratio. 
At high extinctions, A$_{\rm V} > 17$ mag, the Class I/0 to Class II ratio also steeply increases with extinction.
However, this ratio appears flat for the extinction range A$_{\rm V}$ = 8--17 mag.
To investigate this further we look at the contributions to the graph
from the main core and cluster PL07 separately (Fig. 7 b \& c). 
We find that for the main core of the cloud, where most of the star formation is taking place, 
the ratio has a positive monotonic relation with extinction. 
Cluster PL07 region is different; It appears to have a relatively flat relationship between ratio and extinction, with a possible increase at A$_{\rm V}$ = 7--8 mag. 
There is a small offset between the extinction peak and Class I/0 density peak; 
which suggests a recent expulsion of gas from the center of PL07.  
For the cloud as a whole, the mean ratios between the Class I/0 and Class II sources have a small range (0.20--0.60) which may correspond to a narrow range of ages or perhaps evidence for continuous star formation.
\begin{figure}
\includegraphics[width=0.40\textwidth]{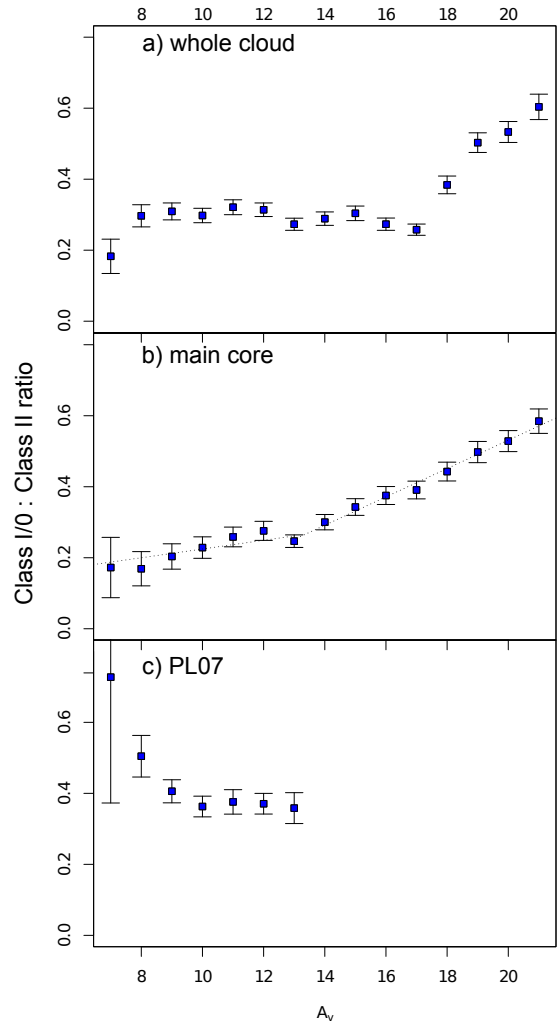}
\caption{Plot of extinction bin versus mean Class I/0 to Class II ratio. 
}
\end{figure}

The relationships between the two ratios and extinction reveal a trend
of decreasing age with increasing extinction. 
This implies that stars may be preferentially forming in the highest extinction 
parts of the molecular cloud.

\section{Discussion}

\subsection{Star formation and column density} 

The Class II to Class III ratio map traces regions with estimated ages
of 0.5--3 Myr, and these regions are spatially coincident with
extinction values of A$_{\rm V}$ = 4--18 mag. 
Using the empirical relationship between disk fraction and cluster age, 
we can fit a power law relation to the Class II to Class III ratio and age. 
Then we can directly study the relation between age and extinction. 
Figure 8 shows a plot of estimated age versus extinction. The plot suggests that the column density of gas decrease exponentially with time above A$_{\rm V}$ = 5 mag, with a half-life, $t_{1/2}$ = 0.4 Myr.
\begin{figure}
\plotone{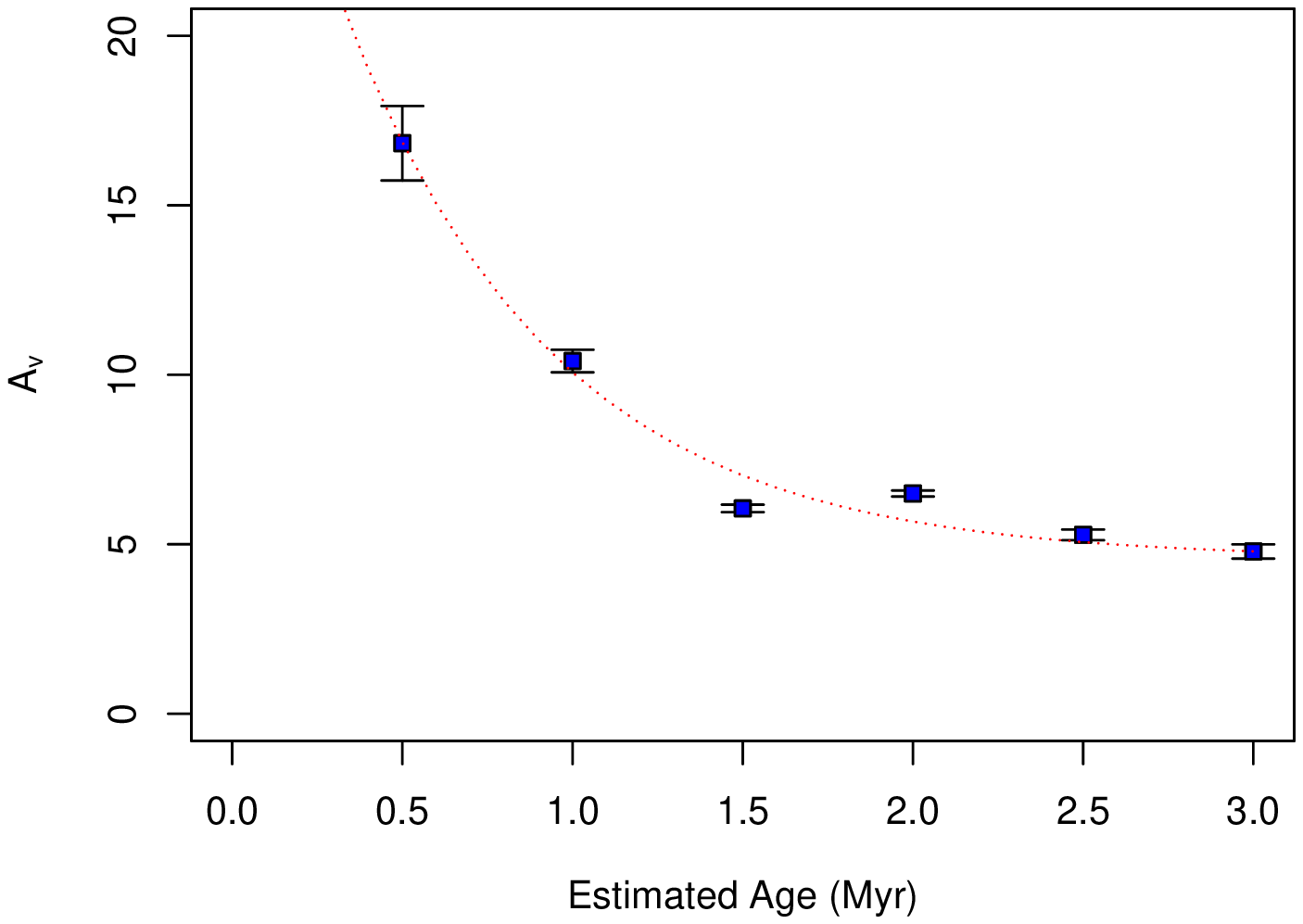}
\caption{Plot of estimated age versus mean extinction. 
The age is estimated from the Class II to Class III ratio.
The red dotted line shows the exponential fit with half-life, $t_{1/2}$ = 0.4 My, above A$_{\rm V}$ = 5 mag.
 }
\end{figure}
We estimate the average rate at which the column density of gas decreases as
\begin{displaymath}
\dot{\Sigma} \sim 10^{-4} \mbox{~M}_{\odot} \mbox{~yr}^{-1} \mbox{~pc}^{-2}
\end{displaymath}
which is over an order of magnitude larger than the star formation
rate measured in nearby molecular clouds
\citep{ladalada2003,evans2009}.  
Thus most of this gas is not removed through formation of stars and is possibly being relocated to other regions of the cloud. 
This is consistent with the study of nearby molecular clouds by \citet{lada2010} which demonstrated that star formation on the scale of few Myr has a negligible effect on the total mass of the cloud.

In the RMC main core, we find a similar relationship between the Class I/0 to Class II ratio and extinction. 
This suggests that star formation occurs preferentially in high extinction regions. 
Additionally, we find that over half of the clustered ($\mu \ge 1.3 \mbox{~stars pc}^{-2}$) Class I/0 sources are found at $A_{V} > 15$ mag and all of them at $A_{V} > 7.5$ mag.
This too is consistent with the study of nearby molecular clouds by \citet{lada2010} that found
an extinction threshold of $A_{V} \sim 7$ mag, above which the star formation rate was proportional to the mass of the cloud measured above that threshold.

\subsection{Cluster Properties}

The relationship between star formation and column density may have consequences
for the formation and evolution of clusters.
Table 2 shows the properties and YSO content of the clusters. 
The stellar content is measured within a 1 parsec (2.14\arcmin) radius of the listed cluster center.

Most of the embedded clusters in the RMC have a Class II to Class III ratio, $R_{II:III}$, between 1 and 2. The weighted mean ratio of all the embedded clusters is $R_{II:III} = 1.2\pm0.2$, suggesting that most of the clusters started forming around the same time. 

There is a group of clusters with ratios suggesting a more recent episode of star formation, having $R_{II:III} > 3.0$. This group includes clusters PL2a, PL02b, PL03b, and REFL09. These clusters however do not appear to have a significant Class I/0 population. This suggests that the star formation in these clusters is more coeval. 

The cluster REFL08 appears to have the most recent episode of star formation. 
This cluster has Class I/0 sources concentrated in dense filamentary
structures seen in extinction. 
It has the highest ratio of Class II to Class III sources which is
consistent with an age less than 1 Myr.  
Our X-ray observations do not cover embedded cluster PL07. 
This cluster has the second highest concentration of young Class I/0 objects. 
However, without knowledge of its Class III content, an age estimate for this region is not possible.  

Although there are some differences between the clusters, the age
spread of these clusters is nonetheless small. 
The ages inferred from the YSO ratios for most of the clusters are between 1 to 3 Myrs. 

\begin{deluxetable*}{lccrrrrrrr}
\tablecaption{Cluster Properties}
\tablehead{
\colhead{Cluster} & \colhead{R.A.} & \colhead{Dec.} & \colhead{I/0} & \colhead{II} & \colhead{III}  & \colhead{II} & \colhead{III}   & \colhead{$R_{II:III}$} & \colhead{$A_{V}$} \\  
\colhead{\null}     &  \colhead{\null} & \colhead{\null} & \colhead{\null} & \colhead{\null} & \colhead{\null} & \colhead{$H \le 14$} & \colhead{$H \le 14$} & \colhead{\null}  & \colhead{\null} 
}
\startdata
PL01   & 97.96 & 4.32 &  5 & 11 & 11 & 10 &  8 &  1.4$\pm$0.7 &  8.1$\pm$1.4\\
PL02   & 98.31 & 4.59 &  2 & 10 &  6 &  7 &  2 &  4.0$\pm$4.0 &  6.4$\pm$2.2\\
PL02b  & 98.31 & 4.53 &  3 & 15 &  5 & 10 &  3 &  3.7$\pm$2.9 &  7.7$\pm$1.0\\
PL03a  & 98.38 & 4.00 &  6 &  4 &  8 &  2 &  6 &  0.5$\pm$0.4 & 10.5$\pm$2.5\\
PL03b  & 98.42 & 4.06 &  2 & 19 &  5 & 12 &  4 &  3.3$\pm$2.2 &  9.9$\pm$1.9\\
PL04a  & 98.53 & 4.42 &  7 & 25 & 14 & 18 & 10 &  1.9$\pm$0.8 & 10.7$\pm$2.4\\
PL04b  & 98.55 & 4.47 &  2 & 32 & 30 & 21 & 17 &  1.3$\pm$0.4 &  7.2$\pm$0.8\\
REFL08 & 98.56 & 4.32 & 18 & 30 &  4 & 19 &  1 &  -           & 17.8$\pm$2.4\\
PL05   & 98.63 & 4.32 &  2 & 32 & 15 & 22 & 11 &  2.1$\pm$0.8 & 11.3$\pm$2.2\\
PL06   & 98.66 & 4.21 &  4 & 10 & 13 &  9 &  9 &  1.1$\pm$0.5 & 10.8$\pm$1.6\\
PL06b  & 98.59 & 4.20 &  0 & 12 & 10 &  7 &  7 &  1.1$\pm$0.7 & 10.2$\pm$1.2\\ 
REFL09 & 98.78 & 3.71 &  5 & 36 &  4 & 16 &  3 &  5.7$\pm$4.4 & 16.6$\pm$2.7\\
PL07   & 98.88 & 3.98 & 12 & 32 & ND & -  & -  &  -           & 10.6$\pm$1.4
\enddata
\tablecomments{The number of sources are counted within 1 pc (2.14\arcmin) of the cluster centers}
\end{deluxetable*}

\subsection{Age gradients in the RMC main core}  

The RMC main core is composed of clusters PL04(a \& b), PL05, and REFL08. 
Each of the clusters is associated with a CO clump identified in
\citet{williams1995} CO survey of the RMC (Table 3). 
These three clusters are characterized by having more YSOs than the other clusters in the cloud, which is consistent with the study by \citet{rom2008} where it is estimated that half of the star formation in the whole cloud happens in this region.

The REFL08 sub-region has the highest density of protostars and is spatially coincident with the gas surface density peak. 
Its ratio of Class I/0 to Class II sources and its dearth of Class III sources suggest the age of this region to be less than 1 Myr. 
\citet{hennemann2010} using Herschel observations found 27 protostars
in this region, 7 of which were classified as very young Class 0
candidates. 
This region appears filamentary in the MIR {\it Spitzer} images with
the main filament running north to south. 
The southern end of the main filament is coincident with the center of
NIR cluster REFL08 and is at the intersection of two smaller
filaments. 
This region appears to be the youngest region of the cloud. 
Cluster PL04 is located north of REFL08, while PL05 is located to the east.
Both clusters, PL04 and PL05, appear to be older with Class II to Class III ratios consistent with ages 1--2 Myrs.

The age gradient across this region as a whole suggests that star
formation began in clusters PL04 and PL05 first, followed by star
formation in REFL08 as a more recent event. 
Based on the YSO content, we estimate the age difference across the region to be about 1 Myr. 
It is possible that star formation feedback from clusters PL04a and PL05 pushed the gas into its present state and thus triggered the formation of REFL08.

Alternatively, the formation of cluster REFL08 may not have been
triggered by PL04 and PL05. The age progression may be a consequence
of the formation of the cloud.  
This scenario is consistent with the numerical simulations of dynamic
molecular cloud formation by \citet{hartmann2012}. 
In these simulations, stars can form from the initial density
fluctuations resulting from turbulence during the formation of a
molecular cloud. 
These stars form before the global gravitational collapse of the cloud leads to a main phase of star formation.

\begin{deluxetable}{ccrr}
\tablecaption{Associated CO clumps}
\tablehead{
\colhead{Cluster} & \colhead{Clump} &\colhead{$\nu$} & \colhead{$\Delta\nu$}
}
\startdata
PL04   &   1 &  15.6 & 2.1 \\ 
PL05   &  19 &  12.2 & 1.7 \\
REFL08 &   6 &  10.8 & 1.9
\enddata
\tablecomments{CO clump data from  \citet{williams1995}}
\end{deluxetable}

\subsection{Star Formation as a function of distance from NGC 2244}

We find that the spatial distribution of young Class I/0 sources
throughout the cloud has little or no correlation to location in the
cloud or distance form NGC 2244. 
The age of NGC 2244 is estimated to be 2--3 Myrs which is consistent
with the ages of the other clusters inferred from the YSO ratios.

Age progressions appear to be limited to small regions, where surface density enhancements lead to more recent star formation episodes. 
Clusters PL02a and PL02b, spatially coincident with the visible edge of the \ion{H}{2} region, 
have Class II to Class III ratios consistent with a recent ($\sim$ 1 Myr) episode of star formation. 
The formation of these clusters may have been triggered by NGC 2244, consistent with the X-ray analysis of \citet{wang2009}.
Our analysis  finds a local age progression surrounding REFL08 which is consistent with the
studies of \citet{rom2008} . 
This progression does not extend through the cloud. 
It appears that any effect the \ion{H}{2} region has on the star formation history is secondary to that of the primordial collapse of the cloud.
Our analysis suggests that star formation started throughout the complex due to a primordial collapse and has progressed through a series of localized episodes of formation closely following each other.

\section{Conclusion}

In our analysis of the different YSO classes in the RMC we find that
they often have different density distributions. 
We use the YSO ratios to study age gradients across the cloud to
better understand how star formation has progressed. 
The relationships between the YSO ratios and extinction suggest that
star formation in the cloud occurs preferentially in high extinction
regions and that the column density of gas rapidly decreases as the region evolves. 
This suggests rapid removal of gas may account for the low star formation
efficiencies observed in molecular clouds.
We find that progressions of star formation appear to be localized 
with a small overall age spread across the cloud,
consistent with star formation starting across the cloud at
roughly the same time. 
The distribution of YSOs in the RMC show little or no effect of NGC
2244 on the relative ages of the clusters except for the possible triggering of clusters PL02a and PL02b.

\acknowledgments

We thank Charles Lada for useful discussions and suggestions in the
writing of this manuscript. 
We also thank Rob Gutermuth for help with the {\it Spitzer} IRAC photometry.
This work is based in part on archival data obtained with the {\it Spitzer
Space Telescope}, which is operated by the Jet Propulsion Laboratory,
California Institute of Technology under a contract with NASA. Support
for this work was provided by an award issued by JPL/Caltech,
a NASA LTSA Grant NNG05GD66G, and a NSF grant AST-1109679.
Partial support for this study provided by the Chandra ACIS Team through the SAO grant SV4-74018.
J.Y. thanks his GSRP advisor, Malcolm Niedner, for encouragement and support.
J.Y. also acknowledges support from a
NASA Goddard Space Flight Center GSRP fellowship NNX10AM07H and a
Space Grant Fellowship from the Florida Space Grant Consortium. 
C.R.-Z. acknowledges support from CONACyT, Mexico, project 152160
and from project UNAM-DGAPA-PAPIIT IA101812. 
J.W. acknowledges support from NASA grant GO1-12030X.



{\it Facilities:} \facility{Spitzer (IRAC,MIPS)}  \facility{CXO (ACIS)}  .



\clearpage

\end{document}